\documentclass[aps,preprint,amssymb,amsmath]{revtex4}
\usepackage{mathrsfs}
\usepackage{graphicx}

\def\half{{\textstyle{\frac{1}{2}}}}
\def\cPT{\mathcal PT}
\def\cP{\mathcal P}
\def\cT{\mathcal T}

\begin{document}

\title{Asymptotic Analysis of the Local Potential Approximation to the Wetterich
Equation}

\author{Carl M Bender$^{a,b}$}\email{cmb@wustl.edu}
\author{Sarben Sarkar$^b$}\email{sarben.sarkar@kcl.ac.uk}
\affiliation{$^a$Department of Physics, Washington University, St.~Louis,
Missouri 63130, USA\\
$^b$Department of Physics, King's College London, London WC2R 2LS, UK}

\begin{abstract}
This paper reports a study of the nonlinear partial differential equation that
arises in the local potential approximation to the Wetterich formulation of the
functional renormalization group equation. A cut-off-dependent shift of the
potential in this partial differential equation is performed. This shift allows
a perturbative asymptotic treatment of the differential equation for large
values of the infrared cut-off. To leading order in perturbation theory the
differential equation becomes a heat equation, where the sign of the diffusion
constant changes as the space-time dimension $D$ passes through $2$. When $D<2$,
one obtains a forward heat equation whose initial-value problem is well-posed.
However, for $D>2$ one obtains a backward heat equation whose initial-value
problem is ill-posed. For the special case $D=1$ the asymptotic series for cubic
and quartic models is extrapolated to the small infrared-cut-off limit by using
Pad\'e techniques. The effective potential thus obtained from the partial
differential equation is then used in a Schr\"odinger-equation setting to study
the stability of the ground state. For cubic potentials it is found that this
Pad\'e procedure distinguishes between a $\cPT$-symmetric $ig\phi^3$ theory and
a conventional Hermitian $g\phi^3$ theory ($g$ real). For an $ig\phi^3$ theory
the effective potential is nonsingular and has a stable ground state but for a
conventional $g\phi^3$ theory the effective potential is singular. For a
conventional Hermitian $g\phi^4$ theory and a $\cPT$-symmetric $-g\phi^4$ theory
($g>0$) the results are similar; the effective potentials in both cases are
nonsingular and possess stable ground states.
\end{abstract}
\maketitle

\section{Introduction}\label{s1}
The renormalization group interpolates between microscopic and macroscopic
scales. In the context of the renormalization group Wilson and Kadanoff
pioneered a nonperturbative approach that led to a reformulation of earlier
perturbative approaches to the role of scale and universality \cite{r1}. The
Wilson-Kadanoff approach involves the construction of effective field theories
in which short-distance fluctuations below a certain length scale are integrated
out. Early work on effective actions that preserve universality may be found in
Ref.~\cite{r2}.

A recent development, known as the {\it functional} renormalization group,
combines the functional formulation of quantum field theory with the Wilsonian
renormalization group \cite{r3,r4,r5}. However, the
functional-renormalization-group equations are intractable. This paper examines
a version of the approach due to Wetterich \cite{r6}, which in the potential
approximation leads to a nonlinear {\it partial differential equation} (PDE)
rather than a functional equation. This simplification enables substantial
progress in understanding the effective potential in arbitrary space-time
dimension.

Our purpose here is to examine the nonlinear PDE due to Wetterich. Our approach
in Sec.~\ref{s2} is unconventional and is based on an asymptotic analysis of the
PDE for large values of the cut-off. This analysis allows us to make general
observations regarding the well-posedness of the leading-order approximation to
the PDE. For $D<2$ the approximation is well-posed but for $D>2$ the
approximation is ill-posed. This observation is based on asymptotic analysis,
but it does provide a clue to understanding the instabilities of numerical
solutions for $D=4$ obtained by making polynomial truncations in effective
potentials \cite{r7}.

Next, we specialize to the quantum-mechanical case $D=1$ in Sec.~\ref{s3} and go
to very high order in our asymptotic approximation. We seek a solution to the
Wetterich PDE as an asymptotic series in powers of the inverse momentum cut-off.
We consider two cases: In the first case the leading term $g\phi^3$ is cubic in
$\phi$. If $g$ is real, the potential is unbounded below and the energy levels
are complex. However, if the leading term is $ig\phi^3$, where $g$ is real, this
is a non-Hermitian but $\cPT$-symmetric potential whose energy levels are {\it
real and bounded below} \cite{r8,r9}. In the second case the leading term $g
\phi^4$ is quartic in $\phi$. If $g>0$, the potential is bounded below and the
energy spectrum is positive. If $g<0$, we interpret the potential as being
$\cPT$ symmetric, and again the energy levels are {\it real and bounded below}
\cite{r8,r9}. The asymptotic expansion respects $\cPT$ symmetry to all orders
and allows us to investigate differences in the renormalization-group flow (for
$D=1$) of the effective actions. We examine the stability of the effective
potential in the limit as the infrared cut-off is extrapolated to zero
by using Pad\'e approximants. This procedure is similar in spirit to the
high-temperature analysis of phase transitions in which Pad\'e approximation is
used to extrapolate from high temperature to finite temperature \cite{r10}.

\leftline{\bf Local potential approximation for the Wetterich functional
equation} 
For completeness we give a brief review of the derivation of the nonlinear PDE.
The Euclidean action for scalar field theories, where the Fourier components of
the field $\Phi$ have momenta of magnitude less than the ultraviolet cut-off
$\Lambda$, is
\begin{equation}
S_\Lambda[\Phi]=\int d^Dx\,\left[\half\partial_\mu\Phi\partial^\mu\Phi+
V_{\Lambda}\left(\Phi\right)\right],
\label{E1}
\end{equation}
where $D$ is the dimension of space-time. If we integrate over Fourier modes
with momenta of magnitude $p$ in the shell $k\leq p\leq\Lambda$, we arrive at an
action that we denote by $S_k[\phi]$, where $\phi$ has Fourier modes with $p\leq
k$. Instead of a sharp infrared cut-off $k$, Wetterich introduced a {\it smooth}
infrared cut-off function $R_k(p^2)$ \cite{r6}, which leads to the concept of an
average action ${\cal S}_k[\phi]$. We consider the renormalization group in the
approximation that the average action ${\cal S}_k [\phi]$ is given by 
\begin{equation}
{\cal S}_k\left[\phi\right]=\int d^Dx\left[\half\partial_\mu\phi\partial^\mu
\phi+V_k(\phi)\right]+\Delta_k[\phi],
\label{E2}
\end{equation}
where $\Delta_k[\phi]=\half\int d^Dp\,\phi_pR_k(p^2)\phi_{-p}$.

A simple smooth cut-off function $R_k(p^2)=(k^2-p^2)\Theta(k^2-p^2)$ that
embodies the properties required for $R_k(p^2)$ was introduced by Litim
\cite{r11}. From the average partition function 
\begin{equation}
Z_k[j]\equiv\exp\left(-W_k[j]\right)\equiv\int{\cal D}\phi\exp\left(-{\cal S}_k
[\phi]-\int_pj_p\phi_{-p}\right)
\label{E3}
\end{equation}
the Legendre transformation gives the averaged effective action \cite{r5}
\begin{equation}
\Gamma_k\left[\phi_c\right]=W_k[j]-\Delta_k\left[\phi_c\right]-\int d^Dx\,
j\phi_c.
\label{E4}
\end{equation}
The renormalization group flow equation is \cite{r5}
\begin{equation}
\partial_k\Gamma_k[\phi_c]=\frac{1}{2}{\rm Tr}\left\{\partial_kR_k\left(\frac{
\delta^2\Gamma_k}{\delta\phi_c(p)\delta\phi_c(q)}+R_k(p^2)\delta(p+q)
\right)^{-1}\right\}.
\label{E5}
\end{equation}

To obtain the Wetterich equation we then make the potential approximation {\it
ansatz} 
\begin{equation}
\Gamma_k\left[\phi_c\right]=\int d^Dx\left(\frac{1}{2}\partial_\mu\phi_c
\partial^\mu\phi_c+U_k\left(\phi_c\right)\right)
\label{E6}
\end{equation}
for the average effective action with Litim's form for $R_k(p)$. The flow
equation then becomes
\begin{equation}
\partial_kU_k\left(\phi_c\right)=\frac{1}{\pi_D}\frac{k^{D+1}}{k^2+U_k''\left(
\phi_c\right)},
\label{E7}
\end{equation}
where $U_k''\left(\phi_c\right)\equiv\frac{d^2}{d\phi_c^2}U_k\left(\phi_c
\right)$, $\pi_D=\frac{D(2\pi)^D}{S_{D-1}}$, and $S_{D-1}=\frac{2\pi^{D/2}}{
\Gamma(D/2)}$ is the surface area of a unit $D$-dimensional sphere. This is the
$D$-dimensional form of the PDE examined in this paper.

We may assume that the equations for $U_k\left(\phi_c\right)$ involve
dimensionless quantities. (If this were not so, we could introduce a mass scale
$M$ to achieve dimensionless variables. For example, for $D=1$ the dimensionless
variables, denoted by a tilde, are $\tilde{\phi}=M^{1/2}\phi$, $\tilde{g}=M^{-3}
g$, $\tilde{k}=M^{-1}k$, and $\tilde{\mu}=M^{-1}\mu$.) The Wetterich equation
(\ref{E7}) can be viewed as being in terms of such dimensionless variables.

To avoid numerical difficulties associated with boundary conditions, in the past
the PDE (\ref{E7}) has been analyzed by approximating $U_k(\phi_c)$ as a finite
series in powers of the field $\phi_c$. This {\it ansatz} leads to a sequence of
coupled nonlinear {\it ordinary} differential equations (see, for example,
Ref.~\cite{r4}). The consistency of such a procedure has not been rigorously
established.

For the quantum-mechanical case ($D=1$) (\ref{E7}) becomes
\begin{equation}
\partial_kU_k\left(\phi_c\right)=\frac{1}{32\pi^2}\frac{k^2}{k^2+U_k''\left(
\phi_c\right)}.
\label{E8}
\end{equation}
Even in this one-dimensional setting, no exact solution to this nonlinear PDE is
known and only numerical solutions have been discussed \cite{r4}.

\section{Large cut-off analysis of the Wetterich Equation}\label{s2}
In this paper we depart from the usual treatment of the Wetterich potential
equation (\ref{E7}) by performing an asymptotic analysis for {\it large} values
of the cut-off $k$. This avoids the arbitrariness involved in restricting the
trajectory of the potential to a truncated function space required in the usual
treatment \cite{r5}. Consequently, our approach also avoids the appearance of
coupled nonlinear ordinary differential equations \cite{r4}. To leading order
the results of this analysis are qualitatively different depending on whether
the space-time dimension $D$ is greater than or less than $2$.

Letting $z=k^{2+D}$ and $U_k(\phi_c)=U(z,\phi)$, we can rewrite (\ref{E7}) as 
\begin{equation}
U_z(z,\phi)=\frac{1}{(2+D)\pi_D}\frac{1}{z^{\frac{2}{2+D}}+U_{\phi\phi}(z,
\phi)},
\label{E9}
\end{equation}
where the subscripts on $U$ indicate partial derivatives. We assume that for
large $z$ we may neglect the $U_{\phi\phi}$ term in the denominator. (The
consistency of this assumption is easy to verify if $D<2$.) To leading order in
our approximation scheme we get
\begin{equation}
U_z(z,\phi)\sim\frac{1}{(2+D)\pi_{D}}z^{-\frac{2}{2+D}}\qquad(z\gg1).
\label{E10 }
\end{equation}
On incorporating a correction $\epsilon$ to this leading behavior
\begin{equation}
U(z,\phi)=\frac{1}{D\pi_D}z^{\frac D{2+D}}+\epsilon(z,\phi),
\label{E11}
\end{equation} 
we get to order $O(\epsilon^2)$
\begin{equation}
\epsilon_z(z,\phi)=-\frac{1}{(D+2)\pi_D}z^{-\frac{4}{2+D}}\epsilon_{\phi\phi}
(z,\phi).
\label{E12}
\end{equation}
On making the further change of variable
\begin{equation}
t=\frac{D+2}{2-D}z^{\frac{D-2}{D+2}},
\label{E13}
\end{equation}
(\ref{E12}) becomes 
\begin{equation}
\epsilon_t(t,\phi)=\frac{1}{(D+2)\pi_D}\epsilon_{\phi\phi}(t,\phi).
\label{E14}
\end{equation}

The variable $t$ in (\ref{E13}) is positive for $D<2$ and negative for $D>2$
and is not defined at $D=2$. Thus, (\ref{E14}) is a conventional diffusion
equation for $D<2$ but is a {\it backward} diffusion equation for $D>2$. The
backward diffusion equation is an inverse problem that is ill-posed \cite{r12}.
The problems
associated with this ill-posedness may be connected with difficulties in solving
(\ref{E7}) numerically when $D=4$ \cite{r3,r7}. 

If $D<2$ in (\ref{E14}), $t$ is positive and $t\to\infty$ as $z\to0$. This
includes the special case $D=1$, which is investigated in Sec.~\ref{s3}.
However, for $D>2$ we can set $\tau=-t$ so that $\tau\to\infty$ as $z\to\infty$.
In terms of the $\tau$ variable we obtain the backwards diffusion equation
\begin{equation}
\partial_\tau\epsilon=-\frac{1}{(D+2)\pi_D}\epsilon_{\phi\phi}.
\label{E15}
\end{equation}

A {\it well-posed} differential-equation problem, as discussed by Hadamard, has
three characteristics: (i) The solution exists; (ii) the solution is unique;
(iii) the solution is stable with respect to small variations of the initial
data. If these three conditions are not met, the differential-equation problem
is {\it ill-posed}.

The $\phi$ variable in the heat diffusion equation (\ref{E14}) can be rescaled
so that the equation appears in the standard form: $\epsilon_t=\epsilon_{\phi
\phi}$. To demonstrate the stability of the forward diffusion equation
$u_t=u_{xx}$ and the instability of the backward diffusion equation $u_t=-u_{xx
}$ we need only consider a small perturbation of the initial data $u(x,0)\to
u(x,0)+\Delta u(x,0)$, which gives rise to a change in the solution at time $t$:
$$u(x,t)\to u(x,t)+\Delta u(x,t).$$
For simplicity, we take $\Delta u(x,t)=\delta e^{at}e^{ibx}$, where for $\delta
\ll1$ this is a small perturbation of the initial data. Then the forward heat
equation gives the dispersion relation $a=-b^2$, so 
\begin{equation}
\Delta u(x,t)=\delta e^{-b^2t}e^{ibx}.
\label{E16}
\end{equation}

Since the heat equation is linear we can use a Fourier representation to
construct {\it any} small perturbation of the initial data as a linear
combination of terms of the form $e^{ibx}$. We can see from (\ref{E16}) that
perturbing the initial data by an amount proportional to $\delta$ gives a
perturbation of the solution for $t>0$ of order $\delta$. Thus, a small change
in the initial data produces a small change in the solution at later times.

The backward heat equation gives the dispersion relation $a=b^2$R, so a small
change in the initial data of the form $u(x,0)\to u(x,0)+\delta e^{ibx}$ gives
rise to a solution that at later times grows exponentially with $t$ like $e^{b^2
t}$. So, small changes in the initial data produce arbitrarily large changes
in the solution and the backwards heat-equation problem is ill-posed.

\section{Series expansions and extrapolation for $D=1$}
\label{s3}
The discussion in Sec.~\ref{s2} shows that instability problems arise when $D>
2$. These problems may just be an artifact of the nonperturbative expansion
techniques proposed in this paper, or perhaps they have a more fundamental
origin. However, there is no difficulty in applying our method in
low-dimensional quantum field theory. In this section we apply our techniques
to the quantum-mechanical case $D=1$. We are especially interested to see what
happens if they are applied to $\cPT$-symmetric quantum-mechanical theories.

On defining $\widehat{U}_k(\phi_c)\equiv U_k(\phi_c)-\frac{k}{\pi}$, (\ref{E8})
becomes
\begin{equation}
\partial_k\widehat{U}_k(\phi_c)=-\frac{\widehat{U}_k''(\phi_c)}{\pi\left[k^2+
\widehat{U}_k''(\phi_c)\right]}.
\label{E17}
\end{equation}
From (\ref{E17}) it is consistent to assume that $\widehat{U}(\phi_c)\to V(
\phi_c)$ and $\widehat{U}''(\phi_c)/k^2\to0$ as $k\to\infty$. (For notational
simplicity we have dropped the suffix $k$ on $\widehat{U}_k$.) The correction to
this asymptotic behavior is $V(\phi_c)+\frac{1}{k}U_1(\phi_c)$ and substituting
in (\ref{E17}) gives
\begin{equation}
U_1(\phi_c)=\frac{1}{\pi}V''(\phi_c).
\label{E18}
\end{equation}
The next corrections are $\widehat{U}(\phi_c)=V(\phi_c)+\frac{1}{k\pi}V''(
\phi_c)+\frac{1}{k^2}U_2(\phi_c)+\frac{1}{k^3}U_3(\phi_c)$ and we get
$$U_2(\phi_c)=\frac{1}{2\pi^2}V^{(4)}(\phi_c)\quad{\rm and}\quad
U_3(\phi_c)=\frac{1}{6\pi^3}V^{(6)}(\phi_c)-\frac{1}{3}[V^{(2)}(\phi_c)]^2,$$
where $V^{(n)}$ is the $n$th derivative of $V$. To formalize this procedure we
let $\delta=\pi/k$ and $\phi=\pi\phi_c$ so that (\ref{E17}) becomes 
\begin{equation}
\frac{\partial}{\partial\delta}\widehat{U}=\frac{\frac{\partial^2}{\partial
\phi^2}\widehat{U}}{1+\delta^2\frac{\partial^2}{\partial\phi^2}\widehat{U}},~{
\rm where}~\widehat{U}=\sum_{n=0}^{\infty}\delta^nU_n(\phi).
\label{E19}
\end{equation}
[A similar analysis of (\ref{E17}) can be performed for $D=2-\eta$ with $\eta$
positive and small: $\widehat{U}(\phi)\sim\sum_na_n(\phi)k^{-n\eta}$. This case
will be studied elsewhere.]

Our perturbation expansion in powers of $\delta$, where $\delta\ll1$ and $k\gg
1$, probes the effective potential at microscopic scales. However, we want to
probe the infrared limit $\delta\to\infty$ of the effective potential. The
generic problem here is obtaining the large-$\delta$ behavior of $\hat{U}$ from
a formal power series that is valid for small $\delta$. There is a parallel in
the theory of phase transitions; there, a physical entity is expanded in inverse
powers of the temperature $T$ \cite{r18} and an extrapolation procedure is used
to find the critical behavior at small $T$.

A powerful extrapolation technique used here to deduce the large-$\delta$
behavior from the small-$\delta$ behavior makes use of Pad\'e approximation
\cite{r19,r20}. Pad\'e approximation consists of converting the formal power
series $\sum_na_n \delta^n$ to a sequence of rational functions 
\begin{equation}
P_M^N(\delta)=\frac{\sum_{n=0}^N A_n\delta^n}{\sum_{n=0}^M B_n\delta^n}.
\label{E20}
\end{equation}
The advantage of constructing $P_M^N(\delta)$ is that in many instances $P_M^N(
\delta)$ is a convergent sequence as $N,M\to\infty$ even when
$\sum_na_n\delta^n$ is a divergent series. 

\subsection{Nonperturbative extrapolation of the series expansion for $D=1$}
Our treatment thus far has been for a general potential. However, we now
specialize to two cases, $U_0(\phi)=g\phi^3$, which is $\cPT$ symmetric for
$g=i$, and $U_0(\phi)=g\phi^4$, which $\cPT$ symmetric for $g$ real and negative
\cite{r8,r16,r17}. We also study the effect of including a mass term of the form
$\mu^2\phi^2$ in $U_0(\phi)$ for both the cubic and quartic cases.

$\cPT$-symmetric field theories are invariant under combined space and
time reflection. The symbol $\cP$ indicates space reflection (parity) and the
symbol $\cT$ indicates time reversal. $\cPT$-symmetric field theories are
complex non-Hermitian deformations of conventional Hermitian field theories in
which the deformation preserves the reality of the eigenvalues. The advantage of
a spectral-reality-preserving deformation is that a new inner product on the
Hilbert space of states can be constructed with respect to which the time
evolution is unitary \cite{r8}. A remarkable feature of $\cPT$-symmetric
theories is that while they may appear to be unstable, the energy spectrum is
real and bounded below, and thus there is actually no instability \cite{r8}.
It is interesting that when one renormalizes a quantum field theory by inserting
counterterms, the effect of doing so can induce non-Hermiticity in the
Hamiltonian for the field theory even though the unrenormalized Hamiltonian is
{\it formally} Hermitian. In some cases, when this happens, the appropriate
approach is to reinterpret the resulting non-Hermitian Hamiltonian as being
$\cPT$ symmetric \cite{r13,r14,r15,r16,r17}. In this paper the
\emph{nonperturbative} functional-renormalization-group equation in the
potential approximation is applied to some elementary $\cPT$-symmetric field
theory models for the first time. This nonperturbative treatment complements our
earlier perturbative study of vacuum instabilities manifest in one-loop
generated effective potentials \cite{r16}.

\subsection{Cubic potentials}
We consider both massless and massive cubic potentials. In the former case
$U_0(\phi)=g\phi^3$ and in the latter case $U_0(\phi)=\mu^2\phi^2+g\phi^3$. 
For the massless case the solution to (\ref{E19}) is 
\begin{equation}
\begin{array}{ll}
U_1(\phi)=6g\phi,&\qquad\quad U_{10}(\phi)=
\textstyle{\frac{1621728}{175}}g^5\phi^3,\\
U_2(\phi)=0,&\qquad\quad
U_{11}(\phi)=\textstyle{\frac{17882208}{1925}}g^5\phi
-\textstyle{\frac{46656}{11}}g^6\phi^6,\\
U_3(\phi)=-12g^2\phi^2,&\qquad\quad 
U_{12}(\phi)=-\textstyle{\frac{30608064}{385}}g^6\phi^4,\\
U_4(\phi)=-6g^2,&\qquad\quad
U_{13}(\phi)=\textstyle{\frac{279936}{13}}g^7\phi^7
-\textstyle{\frac{4537967328}{25025}}g^6\phi^2,\\
U_5(\phi)=\textstyle{\frac{216}{5}}g^3\phi^3,&\qquad\quad
U_{14}(\phi)=\textstyle{\frac{4512754944}{7007}}g^7\phi^5
-\textstyle{\frac{5900745888}{175175}}g^6,\\
U_6(\phi)=\textstyle{\frac{456}{5}}g^3\phi,&\qquad\quad
U_{15}(\phi)=\textstyle{\frac{1439488599552}{525525}}g^7\phi^3
-\textstyle{\frac{1679616}{15}}g^8\phi^8,\\
U_7(\phi)=-\textstyle{\frac{1296}{7}}g^4\phi^4,&\qquad\quad
U_{16}(\phi)=\textstyle{\frac{295324339824}{175175}}g^7\phi
-\textstyle{\frac{25063743168}{5005}}g^8\phi^6,\\
U_8(\phi)=-\textstyle{\frac{34668}{35}}g^4\phi^2,&\qquad\quad
U_{17}(\phi)=\textstyle{\frac{10077696}{17}}g^9\phi^9
-\textstyle{\frac{105103706989824}{2977975}}g^8\phi^4,\\
U_9(\phi)=\textstyle{\frac{7776}{9}}g^5\phi^5
-\textstyle{\frac{89496}{315}}g^4,&\qquad\quad
U_{18}(\phi)=\textstyle{\frac{3212310887424}{85085}}g^9\phi^7
-\textstyle{\frac{101014620538752}{2127125}}g^8\phi^2.
\end{array}
\label{E21}
\end{equation}

These expressions, which are valid for pure imaginary $g$ as well as for real
$g$, are cumbersome but manageable. We stress that there has been no truncation
of the function space on which $\widehat{U}(\phi)$ has support. This contrasts
with the usual approach (for example, see Ref.~\cite{r4}), which requires a
truncation at the onset of the calculation of the renormalization group flow.
This absence of truncation continues to be a feature for the massive case.

The iterative solution to (\ref{E19}) for the massive case is similar to that
for the massless case, but the expressions for $U_{n}(\phi)$ have many more
terms. For example, the coefficient of $\delta^9$ is
\begin{eqnarray}
U_9(\phi)&=&\textstyle{\frac{8}{315}}\big(34020g^5\phi^5+56700g^4\mu^2\phi^4
-11187g^4+37800g^3\mu^4\phi^3\nonumber\\
&&~~+12600g^2\mu^6\phi^2+2100g\mu^8\phi+140\mu^{10}\big),\nonumber
\end{eqnarray}
which in the massless limit $\mu\to0$ reduces to the two-term expression in
(\ref{E21}). We refrain from listing the coefficients explicitly and instead
proceed directly to the large-$\delta$ behavior of the diagonal Pad\'e
approximants. We denote the diagonal Pad\'e approximants in the massless case by
$P_N^{0,N}(\delta)$ and in the massive case by $P_N^N(\delta)$. (This
nonperturbative analysis complements the perturbative analysis given in
Refs.~\cite{r16,r17}.)

Let us examine some low-order Pad\'e approximants in the
limit $\delta\to\infty$. For example,
$$\lim_{\delta\to\infty}P_2^2(\delta)=\frac{18g^3\phi^5+3g^2\left(10\mu^2\phi^4
-9\right)+14g\mu^4\phi^3+2\mu^6\phi^2}{2\left(3g\phi+\mu^2\right)^2}.$$
[As a check, when $\mu=0$ this expression agrees with the corresponding
expression $g\phi^3-\frac{3}{2\phi^2}$ for $P_2^{0,2}(\delta)$.] For $N=3$ we
obtain
$$\lim_{\delta\to\infty}P_3^{0,3}(\phi)=\frac{g\phi^3\left(736g\phi^5+1705
\right)}{25-800g\phi^5}\qquad{\rm and}\qquad P_3^3(\phi)=-u_3/d_3,$$
where
\begin{eqnarray}
u_3&=&1609632g^8\phi^8+729g^7\phi^3\left(5888\mu^2\phi^4+5115\right)
+243g^6\mu^2\phi^2\left(23008\mu^2\phi^4+15345\right)\nonumber\\
&&~~+1296g^5\mu^4\phi\left(3376\mu^2\phi^4+945\right)+15120g^4\mu^6\left(142
\mu^2\phi^4+9\right)+658944g^3\mu^{10}\phi^3\nonumber\\
&&~~+121824g^2\mu^{12}\phi^2+12288g\mu^{14}\phi+512\mu^{16},\nonumber
\end{eqnarray}
$$d_3=225g^2[7776g^5\phi^5+81g^4(160\mu^2\phi^4-3)+8640g^3\mu^4
\phi^3+2880g^2\mu^6\phi^2+480g\mu^8\phi+32\mu^{10}].$$
[As a check, we have $\lim_{\mu\to0}P_3^3(\phi)=P_3^{0,3}(\phi)$.]

\begin{figure}[h!]
\begin{center}
\includegraphics[scale=0.27]{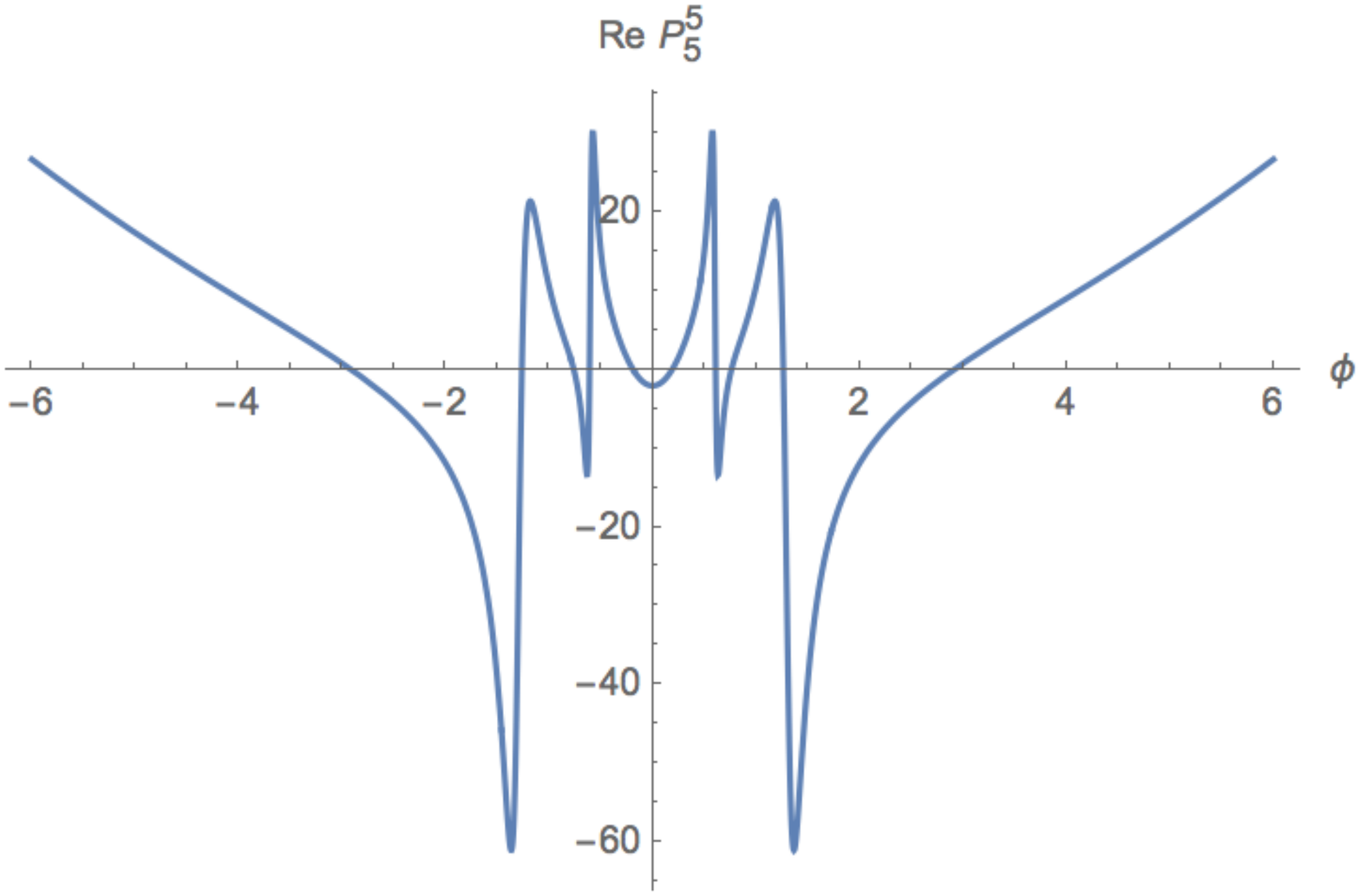}\hspace{.1in}
\includegraphics[scale=0.27]{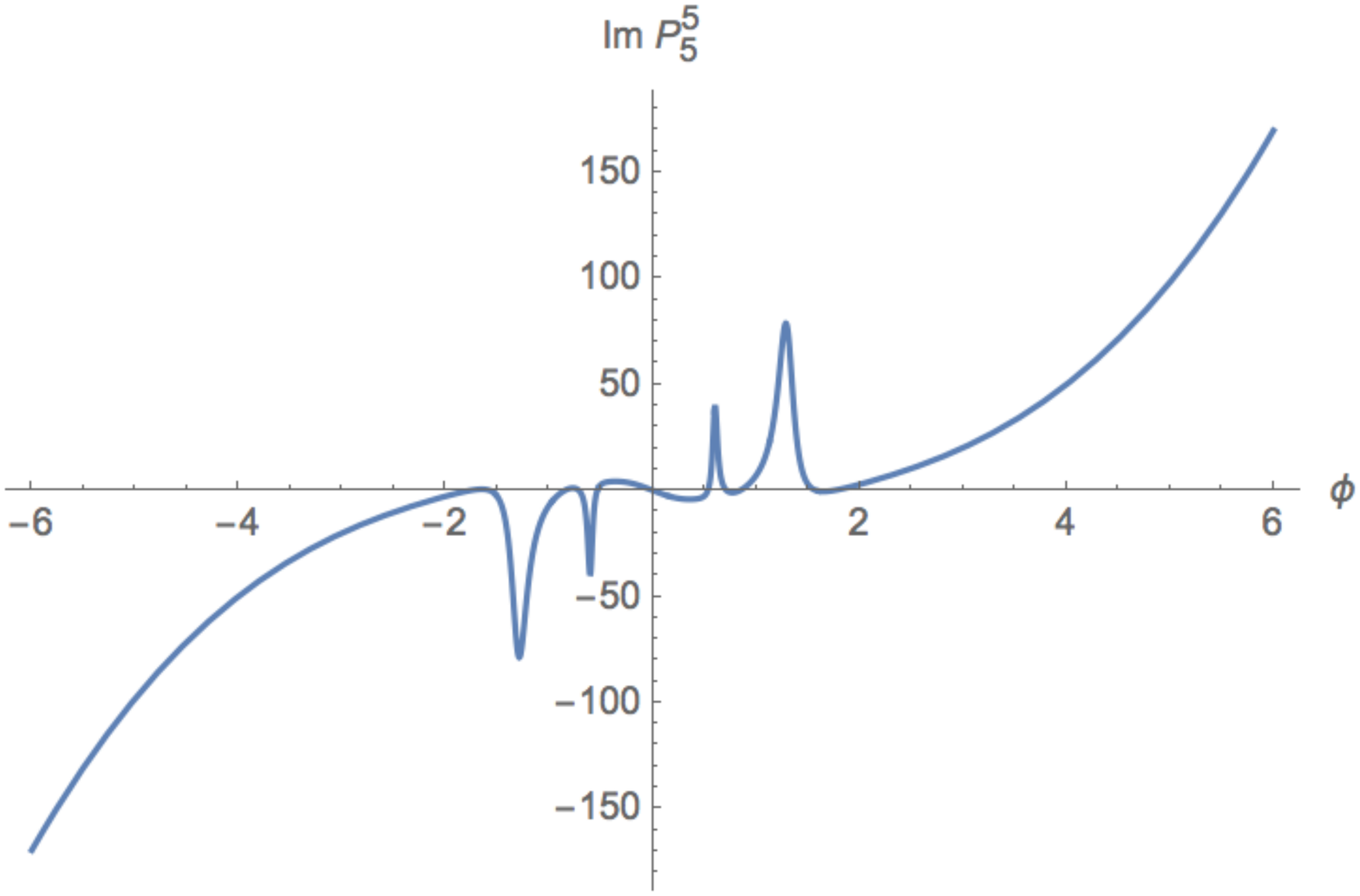}
\end{center}
\caption{Effective potential flow for massive $i\phi^3$ with the mass parameter
$\mu=1$. Shown is a plot of the real and imaginary parts of the $P_5^5$
approximant plotted as functions of real $\phi$. Observe that there are no
poles. The real part of the potential is right-side-up, so there is no
instability. Furthermore, apart from small fluctuations near the origin, the
imaginary part of the potential behaves like $i\phi^3$ for large $|\phi|$.}
\label{f1}
\end{figure}

\begin{figure}[ht]
\begin{center}
\includegraphics[scale=0.27]{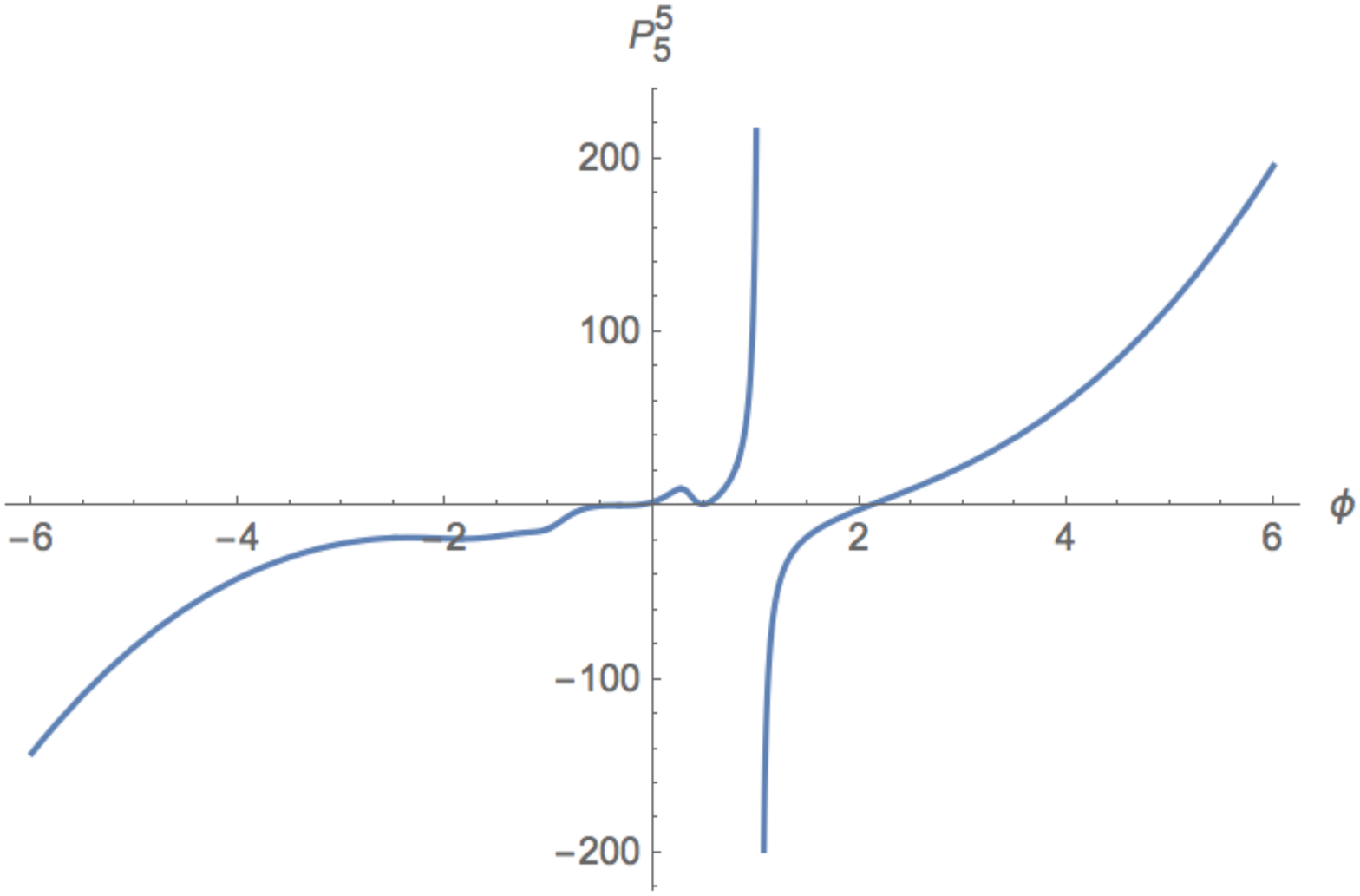}\hspace{.1in}
\includegraphics[scale=0.27]{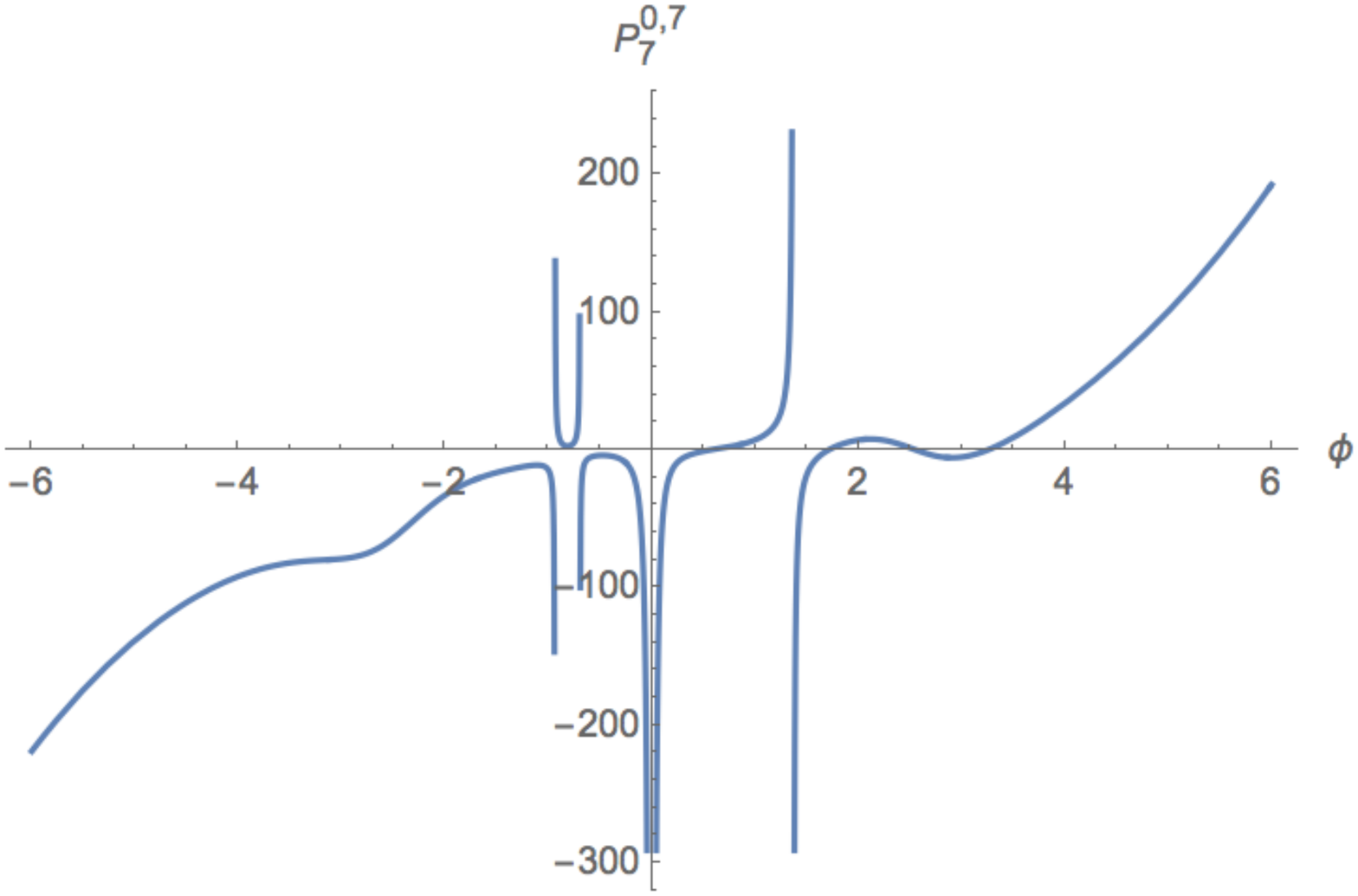}
\end{center}
\caption{Pad\'e $P_5^5$ approximation for the effective potential flow for
massive $\phi^3$ and $P_7^7$ for massless $\phi^3$ potentials. These potentials
rise on the right but fall on the left. Thus, this theory is unstable and
nonconfining. Note also that there are poles on the $\phi$ axis. There are no
such poles for the $\cPT$-symmetric $i\phi^3$ theory, as we can see in
Fig.~\ref{f1}.} 
\label{f2}
\end{figure}

Note that for the massless case $P_2^{0,2}$ has a double pole for both real and
imaginary $g$. In fact, the same is true for $P_n^{0,n}$ for even $n$. This pole
is an artifact of the Pad\'e approximation and is not present when $n$ is odd,
so we will only consider the behavior of these approximants for odd $n$. In
general, the odd-$n$ diagonal Pad\'e approximants have no singularities at all
on the real-$\phi$ axis when $g$ is imaginary (see Fig~\ref{f1} for the case
$n=5$) but singularities occur for the case of real $g$ (see Fig.~\ref{f2}).
These findings indicate that the $\cPT$-symmetric effective potential is well
behaved in the infrared limit (see Fig.~\ref{f1}). From the expressions for the
diagonal Pad\'e approximants we see that for large $|\phi|$, the leading
behavior of the imaginary part of the effective potential is exactly $i\phi^3$.
Consequently the $\cPT$ nature of the interaction is preserved under
renormalization (see the discussion of the effective potential obtained from the
perturbative renormalization group in \cite{r16,r17}).

The situation is not so nice for the conventional $g\phi^3$ theory ($g$ real).
In Fig.~\ref{f2} we plot the Pad\'e approximations $P_5^5$ for the effective
potential flow for massive $\phi^3$ and $P_7^7$ for massless $\phi^3$. These
plots indicate that poles appear on the real-$\phi$ axis. We emphasize that
there are no such poles for the $\cPT$-symmetric $i\phi^3$ theory. Furthermore,
these potentials rise for positive $\phi$ and fall for negative $\phi$, which
indicates that the theory is unstable.

\subsection{Pad\'e approximants for $\cPT$-symmetric quartic theories}
We consider next the case of massless and massive $\cPT$-symmetric quartic
theories; that is, theories with potentials of the form $-g\phi^4$ and $\mu^2
\phi^2-g\phi^4$. Potentials such as these are obtained from the general
one-parameter class of potentials of the form $\phi^2(i\phi)^\epsilon$ and
$\mu^2\phi^2+\phi^2(i\phi)^\varepsilon$, where $\varepsilon\geq0$. These
potentials interpolate between the quadratic and the cubic and quartic
$\cPT$-symmetric potentials considered in this paper: For $\varepsilon=2$ we
obtain the negative-quartic potential and for $\varepsilon=1$ we obtain the
imaginary cubic potential \cite{r8}. The functional integral representing the
partition function for a $\cPT$-symmetric $-\phi^4$ theory must be defined
inside a pair of Stokes sectors of opening angle $\frac{\pi}{4}$ that are
centered about the angles $\frac{\pi}{4}$ below the positive-real-$\phi$ axis
and the negative-real-$\phi$.

To be precise, as soon as $\varepsilon>0$, a logarithmic branch point appears at
the origin $\phi=0$. We choose the branch cut to run up the imaginary-$\phi$
axis from $\phi=0$ to $\phi=i\infty$. In this cut plane the integrand of the
functional integral is single valued. The Stokes sectors in which the functional
integral converges rotate downward into the complex-$\phi$ plane and the opening
angles of the sectors decrease as $\varepsilon$ increases. (There are, in fact,
many sectors inside of which the functional integral converges, but we choose to
continue the field theory off the conventional free field theory (the harmonic
oscillator for the case $D=1$), which is obtained when we set $\varepsilon=0$.
For $\varepsilon>0$ the center lines of the left and right sectors lie at the
angles
\begin{equation}
\label{E22}
\theta_{\rm left}=-\pi+\frac{\pi\varepsilon}{2\varepsilon+4}\quad{\rm and}\quad
\theta_{\rm right}=-\frac{\pi\varepsilon}{2\varepsilon+4}.
\end{equation}
The opening angle of each sector is $\Delta=\pi/(2+\varepsilon)$.

\begin{figure}[ht]
\begin{center}
\includegraphics[scale=0.27]{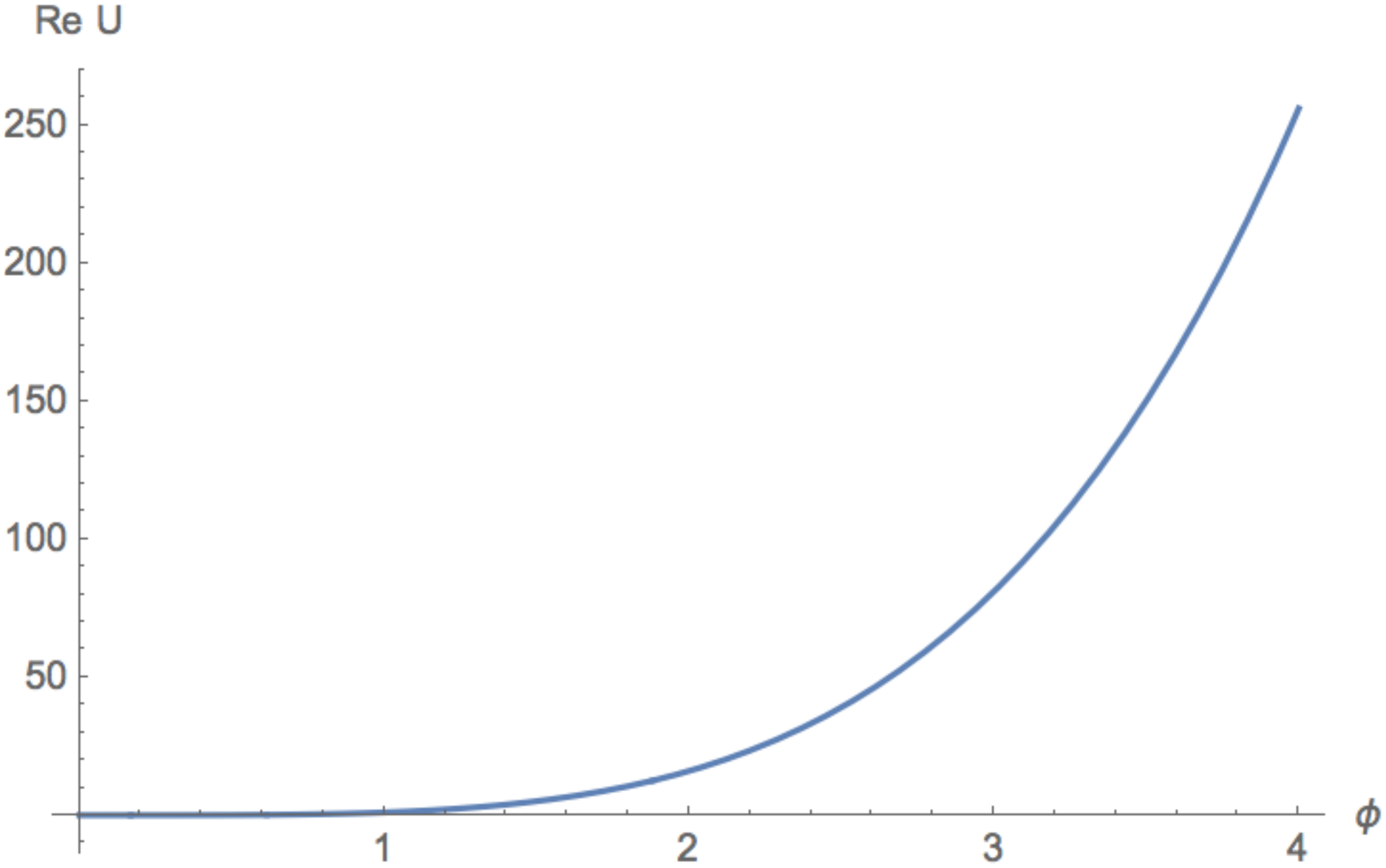}\hspace{.1in}
\includegraphics[scale=0.27]{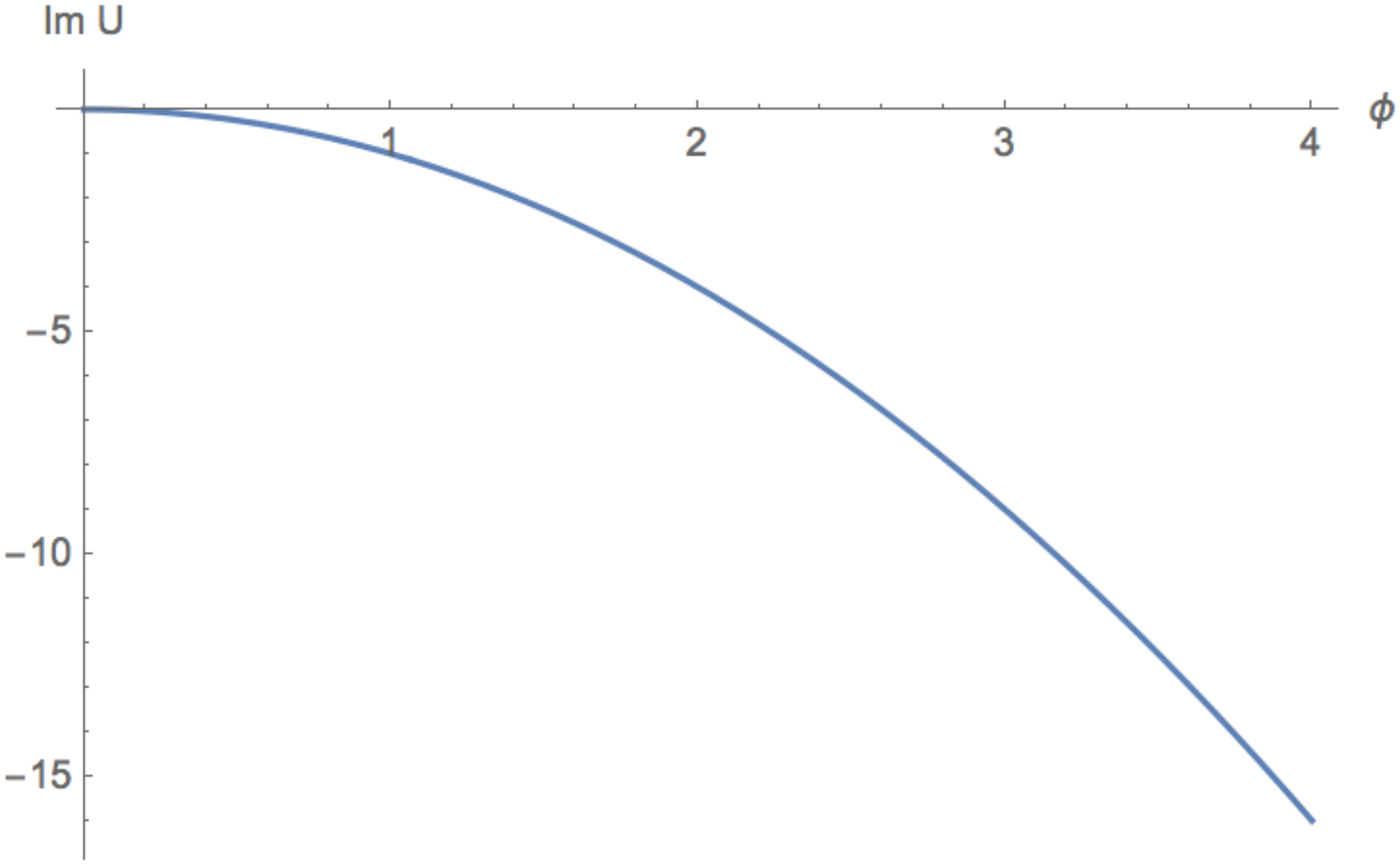}
\end{center}
\caption{Real and imaginary parts of the quartic potential $-\phi^4+\phi^2$
along the center line of the right Stokes sector. The functional integral
converges because the real part of the potential rises along this line.}
\label{f3}
\end{figure}

The functional integral may be performed along any contour in the complex-$\phi$
plane so long as the endpoints of the contour lie inside the left and right
sectors. When $0\leq\varepsilon<1$ the sectors contain the real axis. However,
if $\varepsilon\geq1$, the sectors rotate below the real-$\phi$ axis. In
Fig.~\ref{f3} we plot the massive quartic potential $-\phi^4+\phi^2$ along the
center line in the right Stokes sector. Note that the real part of the potential
{\it rises} along this; this is why the functional integral for the partition
function converges for the upside-down quartic potential.

When we compute the Pad\'e approximants for the $\cPT$-symmetric
(negative-quartic-potential) case, we find that there are poles at the origin
for $P_2^2$, $P_5^5$, $P_8^8$, and so on. We interpret this problem as an
artifact of the Pad\'e procedure, and we disregard every third Pad\'e
approximant. (We observe an interesting pattern here: Every second Pad\'e must
be discarded for the cubic potential and every third Pad\'e must be disregarded
for the quartic potential.) The large-$|\phi|$ behavior of the Pad\'e
approximants has the form $\alpha g\phi^4$. From $P_3^3(\phi)$ we find that
$\alpha=33/97=0.340206\ldots$; from $P_4^4(\phi)$ we find that $\alpha=1$; from
$P_6^6(\phi)$ we find that $\alpha=1$; from $P_7^7(\phi)$ we find that $\alpha=
0.977905\ldots$, and so on. Thus, the even-even diagonal Pad\'e approximants
yield the value $\alpha=1$ while the odd-odd diagonal Pad\'e approximants yield
positive values for $\alpha$ that rapidly and monotonically approach $1$ from
below. 

\begin{figure}[ht]
\begin{center}
\includegraphics[scale=0.27]{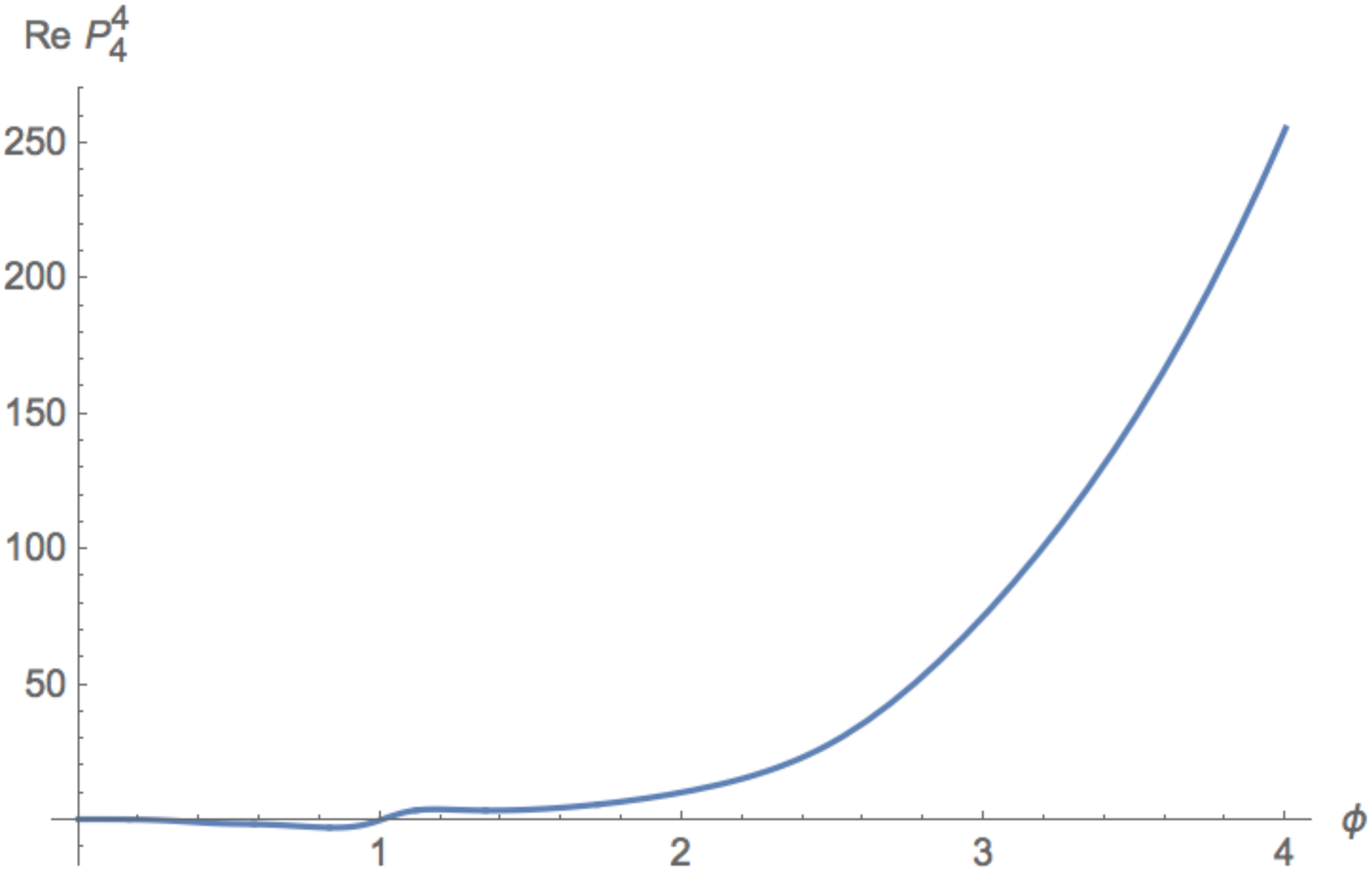}\hspace{.1in}
\includegraphics[scale=0.27]{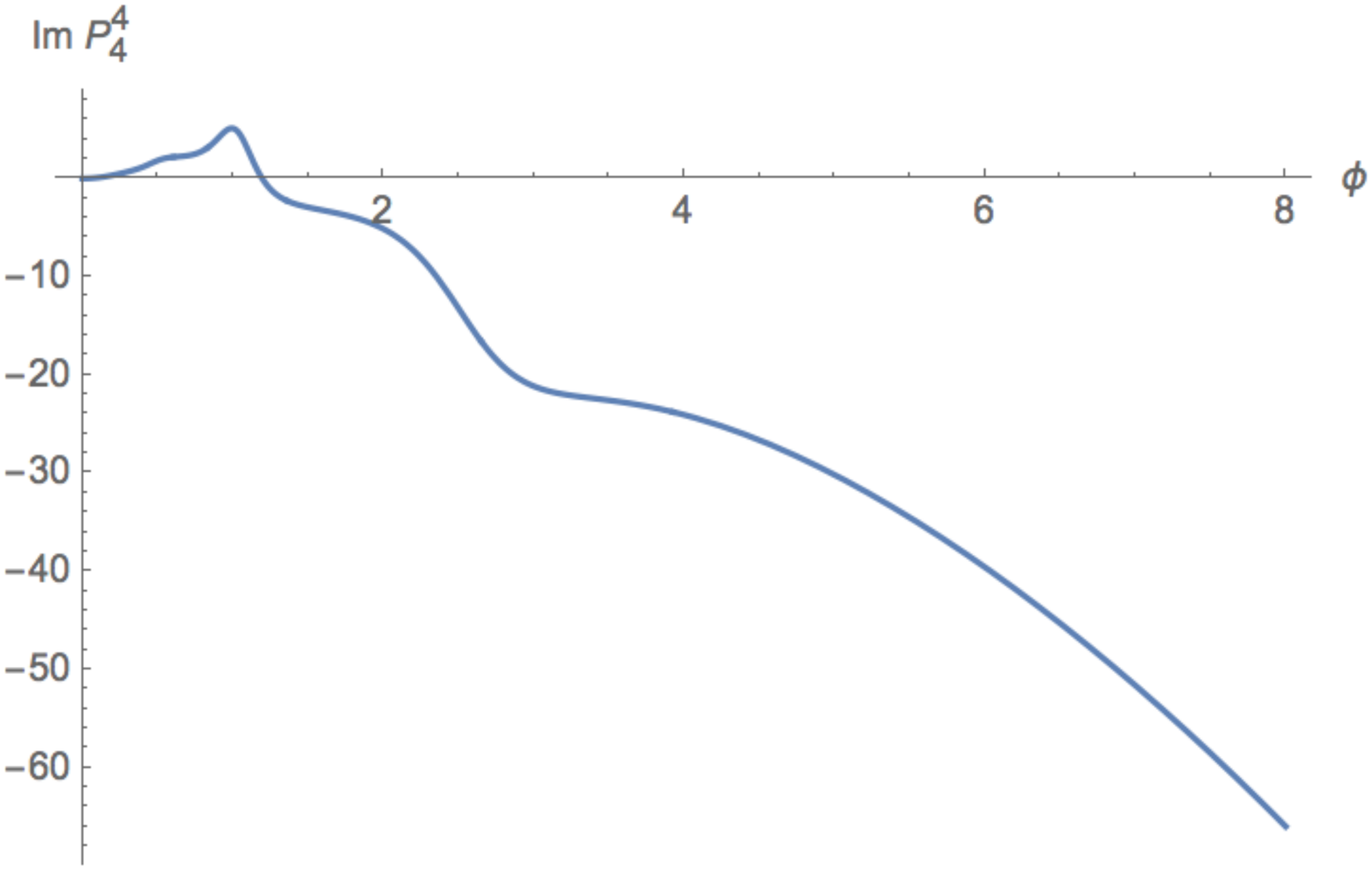}
\end{center}
\caption{Effective potential obtained from $P_4^4(\phi)$ plotted along the
center line of the right Stokes sector for the negative-quartic massive theory
$U_0(\phi)=\mu^2\phi^2-g\phi^4$ with $g=1$ and $\mu=1$. Note the good agreement
with the plot of the potential in Fig.~\ref{f3}.}
\label{f4}
\end{figure}

\begin{figure}[ht]
\begin{center}
\includegraphics[scale=0.27]{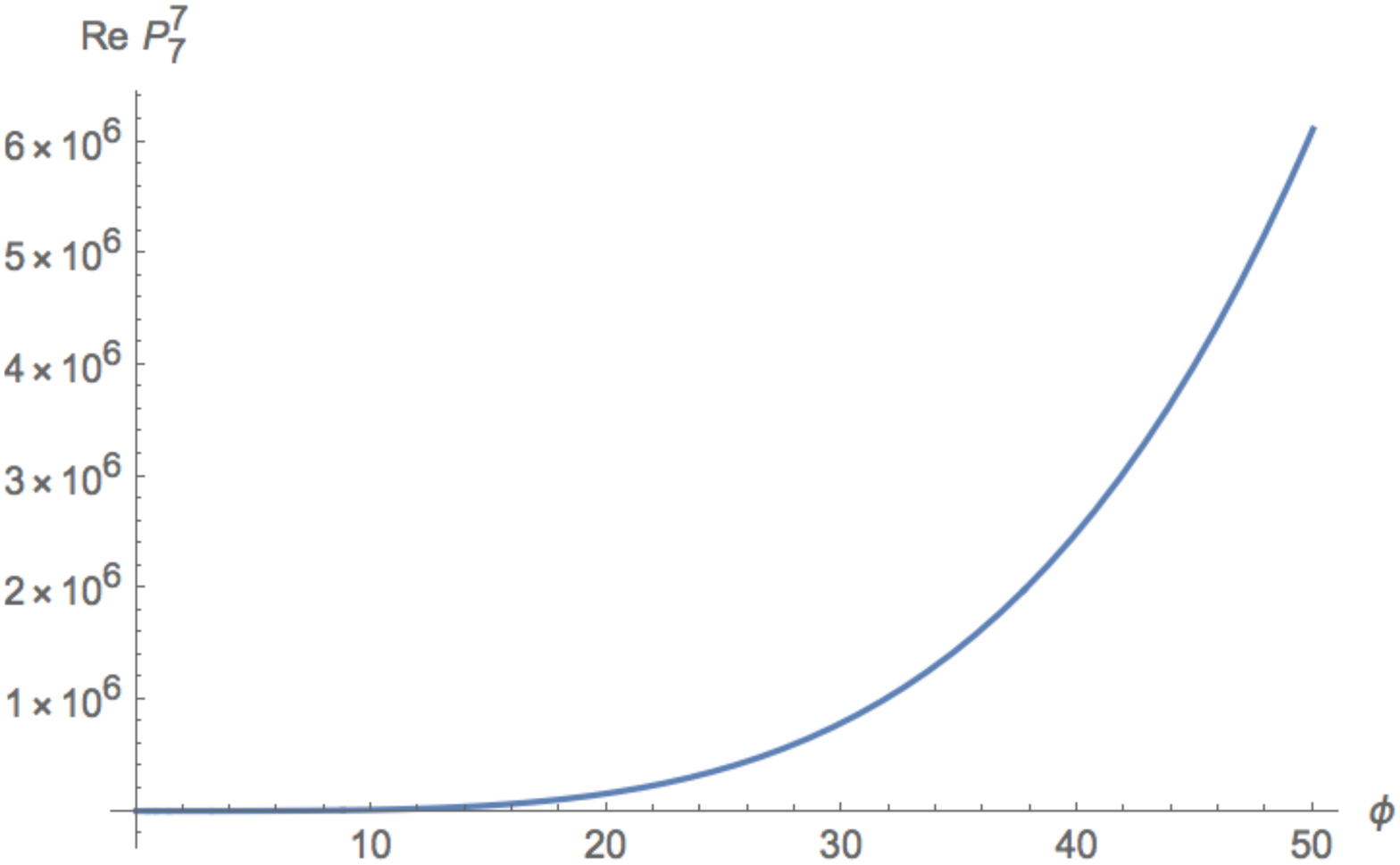}\hspace{.1in}
\includegraphics[scale=0.27]{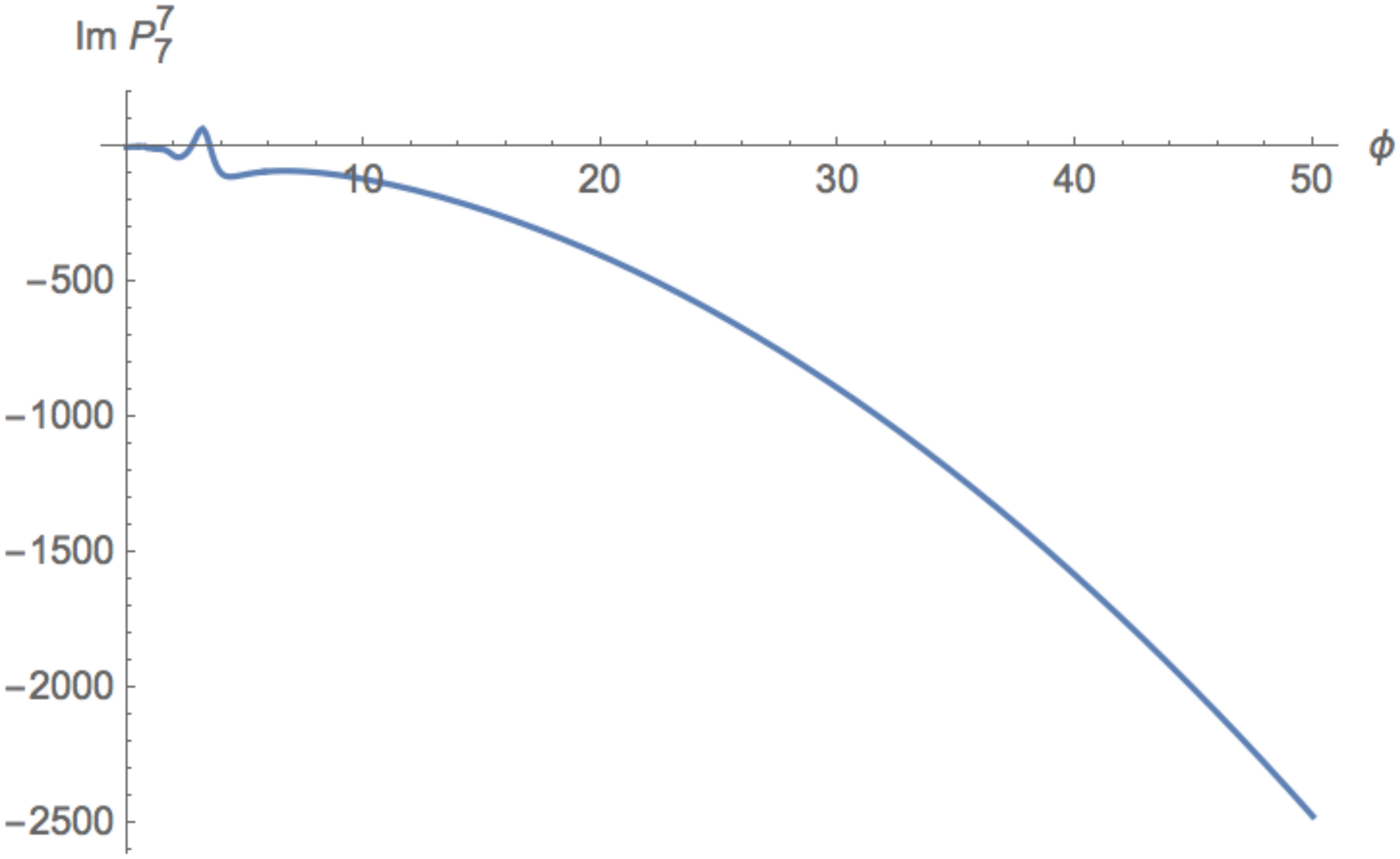}
\end{center}
\caption{Effective potential obtained from $P_7^7(\phi)$ plotted along the
center line of the right Stokes sector for the negative-quartic massive theory
with $g=1$ and $\mu=1$. Like the plot in Fig.~\ref{f4}, there is impressive
agreement with the plot of the potential in Fig.~\ref{f3}.}
\label{f5}
\end{figure}

In Figs.~\ref{f4} and \ref{f5} we plot the real and imaginary parts of the
effective potential at long scales that flows from massive potential $U_0(\phi)=
\mu^2\phi^2-g\phi^4$. The Pad\'e approximants $P_4^4(\phi)$ and $P_7^7(\phi)$
are plotted  along the center line of the right Stokes sector. Note that for
large $|\phi|$ the effective potentials that are obtained in the Pad\'e
approximation strongly resemble the plot of the potential in Fig.~\ref{f3}.
The results for the massless case are equally impressive. These Figures provide
strong support for the perturbative treatment proposed in this paper and give
evidence that $\cPT$-symmetric potentials give rise to valid and consistent
quantum field theories.

\section{Discussion}\label{s4}
The functional renormalization group in the potential approximation provides a
seemingly simple nonlinear partial differential equation. However, we have shown
using asymptotic analysis that the space dimension $D$ determines whether an
initial-value problem well-posed or ill-posed; if $D<2$, the problem is
well-posed and if $D>2$ the problem is ill-posed. For D=1 we have used
asymptotic analysis to investigate $\cPT$-symmetric potentials. Under the
Wetterich flow to the infrared we find that $\cPT$ symmetry is preserved. Vacuum
stability inherent in $\cPT$ symmetric theories persists under renormalization.

We have found that under the functional renormalization group transformation the
effective action of $\cPT$-symmetric theories flows to the infrared in a
$\cPT$-symmetric fashion. This indicates that our earlier discussions of $\cPT$
symmetry and vacuum stability \cite{r16,r17} are relevant in a nonperturbative
setting. The role of renormalization in selecting $\cPT$-symmetric boundary
conditions for field theories is an important question that requires further
study.

\acknowledgments
We thank J.~Alexandre and I.~Saltas for helpful discussions.

\end{document}